\documentclass[nohyper,12pt,letterpaper]{JHEP}
\usepackage{amsfonts,amssymb}

\date{August 2, 2000}

\def\be{\begin{equation}}
\def\ee{\end{equation}}
\def\bear{\begin{eqnarray}}
\def\eear{\end{eqnarray}}
\def\nn{\nonumber}

\newcommand{\px}[1]{{\partial_{#1}}}
\newcommand{\qx}[1]{{\partial^{#1}}}

\newcommand{\rep}[1]{{{\bf {#1}}}}      
\newcommand{\tr}[1]{{\mbox{tr}\{{#1}\}}}          
\newcommand{\com}[2]{{\lbrack {#1},{#2}\rbrack}}  

\def\BZ{\mathbb{Z}}

\newcommand{\MR}[1]{{\mathbb{R}^{#1}}}               
\newcommand{\MS}[1]{{{\bf S}^{#1}}}               
\newcommand{\MT}[1]{{{\bf T}^{#1}}}               

\newcommand{\SUSY}[1]{{{\cal N}= {#1}}}           

\def\a{{\alpha}}

\def\u{{\mu}}
\def\v{{\nu}}

\def\Th{{\Theta}}

\def\btau{{\overline{\tau}}}

\newcommand{\batop}[2]{{\left\{\begin{array}{c}
      {#1}\\ {#2}\\ \end{array}\right\}}}

\title{Dipoles, Twists and Noncommutative Gauge Theory}
\author{Aaron Bergman and Ori J. Ganor\\
Department of Physics, Jadwin Hall \\
Princeton University \\
NJ 08544, USA
}

\abstract{
T-duality of gauge theories on a noncommutative
$T^d$ can be extended to include fields with twisted boundary
conditions.
The resulting T-dual theories contain novel nonlocal fields.
These fields represent dipoles of constant magnitude.
Several unique properties of field theories on noncommutative
spaces have simpler counterparts in the dipole-theories.
}

\received{???????? ?st, 2000} \accepted{???????? ?th, 2000}

\preprint{\hepth{0008030}; PUPT-1946}

\begin{document}


\section{Introduction}\label{intro}
Gauge theories on a noncommutative $\MT{d}$ possess
a T-duality that acts on the metric $G_{ij}$ and anti-symmetric
2-form $\Th_{ij}$ \cite{CDS}-\cite{SWNCG}.

In this paper we will explore the action of T-duality
on noncommutative field theories  with twisted
boundary conditions. Suppose we take a scalar field $\Phi(x)$
with boundary conditions
$\Phi(x_1 +2\pi n_1,\dots,x_d+2\pi n_d) =
e^{i(n_1 \a_1+\cdots +n_d\a_d)}\Phi(x_1,\dots,x_d)$,
where $(x_1,\dots,x_d)$ are coordinates on $\MT{d}$ with period
$2\pi$ and $(\a_1,\dots,\a_d)$ are ``twists.''
The question is: what happens after T-duality?

We will show that the T-dual of such a theory contains nonlocal
fields that behave as constant dipoles, even when the noncommutativity
is turned off. The dipole-vector and the twists, $\a_i$, together
form a $2d$-dimensional representation of the $SO(d,d,\BZ)$
T-duality group.

The paper is organized as follows.
In section (\ref{revtdl}) we review the proof of T-duality in
noncommutative gauge theories.
In section (\ref{twists}) we extend T-duality to act on the twists.
We define the dipole theories and show that the twists, $\a_i$,
and the dipole-vectors transform into each other under T-duality.
In section (\ref{prop}) we explore the properties of
the dipole theories.
These have several features that are reminiscent
of noncommutative field theories, although these features seem
to  have a much simpler version in the dipole theories.
For example, we can define a modified product of fields, and we describe
the analog of the ``Seiberg-Witten map'' \cite{SWNCG} to local variables.
We also show that when compactified
on $\MS{1}$, the dipole theories reduce
to ordinary quiver theories \cite{DouMoo} when the dipole-vector
is a rational fraction of the circumference of $\MS{1}$.

Before we proceed, let us note that
dipoles, in the context of noncommutativity, are also discussed
in \cite{BigSus} and in an upcoming paper \cite{BerkTA}.

\section{Review of T-duality in Noncommutative Gauge Theories}
\label{revtdl}
Unlike the commutative theory, noncommutative Yang-Mills theory exhibits
the T-duality of string theory. T-duality of non-commutative tori was
first investigated in \cite{CDS} in the context of compactifications of Matrix
theory. In the context of noncommutative geometry, T-duality is
implemented as a Morita equivalence between the C*-algebras that are the
noncommutative tori \cite{ASchw,RSchw}. In this language, vector
bundles correspond to projective modules and the Morita equivalence is
given by a bimodule which allows us to map modules over one torus to
modules over the other \cite{RW}.  T-duality of noncommutative gauge
theories is also reviewed in
\cite{AsNeSc}-\cite{SWNCG}.

These considerations are somewhat abstract, however, so we will now give
an explicit relation between adjoint fields on non-commutative tori. This
will lead to a construction for the transformation of the covariant
derivative and the gauge connection.

The theories we work with are $U(n)$ gauge theories with $m$ units of
electric flux. We will show that any such theory is dual to a
$U(\gcd(n,m))$ theory with no flux and a different noncommutativity
parameter. The statement of T-duality is that any pair of
T-dual theories correspond
to the same zero flux theory. Our presentation of the T-duality of
the fields is a slight generalization of the construction in
\cite{MZ}. We will work
on $T^2$ with noncommutativity parameter
$\theta$. This means that we work with the algebra of functions on the
torus subject to the following relation:
\begin{equation}
	[x,y] = 2\pi i R^{2} \theta
\end{equation}

\noindent where $(2\pi R)^{2}$ is the area of the torus.

We collect some useful facts that follow from this relation and the
Baker-Campbell-Hausdorff formula:
\begin{eqnarray}
e^A B e^{-A} &=& e^{Ad(A)} B \label{eabema},\\
\log{e^A e^B} &=& A + B + \frac{1}{2}[A,B] + \ldots, \label{BCH}\\
\left[x,f(x,y)\right] &=& 2{\pi}iR^{2}\theta\partial_y{f(x,y)},\nn\\
\left[y,f(x,y)\right] &=& -2{\pi}iR^{2}\theta\partial_x{f(x,y)},
\label{com2der}
\end{eqnarray}

\noindent {}From equation (\ref{eabema}), we obtain the useful relation
\begin{equation}
\label{trans}
e^{ax}f(x,y)e^{-ax} = e^{2\pi iaR^{2}\theta\partial_{y}}f(x,y) = f(x,y+2\pi
iR^{2}a\theta)
\end{equation}

Here and up until the end of section (\ref{twists}), a product
indicates the noncommutative $\star$-product.

A bundle over the torus with nonzero flux is given by a pair of
transition functions such that an adjoint section transforms as:
\begin{eqnarray}
\label{bdrycond}
\Psi(x+2\pi R,y) = \Omega_{1}(x,y) \Psi(x,y)
           \Omega_{1}(x,y)^{-1} \nn\\
\Psi(x,y+2\pi R) = \Omega_{2}(x,y) \Psi(x,y) \Omega_{2}(x,y)^{-1}
\end{eqnarray}

\noindent A consistent choice of transition functions is:
\begin{equation}
	\Omega_{1} = e^{imy/nR}U \qquad \Omega_{2} = V
\end{equation}

\noindent where $U$ and $V$ are matrices satisfying
\begin{equation}
UV = e^{2\pi im/n} VU \qquad U^{n} = V^{n} = 1
\end{equation}

Let $\gcd(m,n) = \nu$, $\tilde{m} = m/\nu$, and $\tilde{n} = n/\nu$.
We define the following $\tilde{n} \times \tilde{n}$ matrices:
\begin{equation}
	u_{kl} = e^{{2\pi}ikm/n} \delta_{kl} \qquad
v_{kl} = \delta_{k+1,l} \qquad k,l \in \BZ/\tilde{n}\BZ
\end{equation}

\noindent Our choice for $U$ and $V$ will be the $n \times n$ matrices
that have $\nu$ copies of $u$ and $v$ along the diagonal. In the case
of $\nu = 1$, these are the matrices of \cite{MZ}.

We can now put the field $\Psi$ into a standard form. Following
\cite{MZ}, we note
\begin{equation}
        \Psi(x + 2\pi R\tilde{n},y) =
\Omega_{1}^{\tilde{n}}\Psi(x,y)\Omega_{1}^{-\tilde{n}} =
         \Psi(x + 2\pi R\theta \tilde{m},y)
\end{equation}

\noindent
Therefore, we have the following periodicity conditions:
\be
	\Psi(x+2\pi R(\tilde{n}- \tilde{m}\theta),y) = \Psi(x,y),\qquad
	\Psi(x,y+2\pi R\tilde{n}) = \Psi(x,y)
\ee

\noindent and we can do a Fourier expansion:
\begin{equation}
\Psi(x,y) = \sum_{s,t\in\BZ} e^{isx/(\tilde{n} - \tilde{m}\theta)}
e^{-ity/\tilde{n}}
\Psi_{s,t}
\end{equation}

$\Psi_{s,t}$ is a $n \times n$ matrix which we treat as a
matrix of $\nu \times \nu$ blocks. Thus,
we have the $\tilde{n} \times \tilde{n}$ matrix
$\Psi^{f,g}_{s,t}$ with $f,g \in \BZ_{\nu}$. We expand
this matrix in terms of the $u$ and $v$ matrices:
\begin{equation}
\Psi^{f,g}_{s,t} = \sum_{i,j \in \BZ/\tilde{n}\BZ}  c^{f,g}_{s,t,i,j}
v^{i} u^{j}
\end{equation}

In \cite{MZ} it is shown that, once we impose the boundary
conditions (\ref{bdrycond}), only one term is nonzero in this sum.
Their final result, which does not depend on the presence of our $f$
and $g$ indices, is the expansion:
\begin{equation}
        \label{stdform}
        \Psi^{f,g} = \sum_{s,t\in\BZ} c^{f,g}_{s,t} Z_{1}^{s} Z_{2}^{-t}
\end{equation}

\noindent where $a\tilde{n}-b\tilde{m} = 1$,
$c^{f,g}_{s,t}$ is the appropriate nonzero element out of
$c^{f,g}_{s,t,i,j}$ and we have:
\begin{equation}
Z_{1} = e^{ix/R(\tilde{n}-\tilde{m}\theta)} v^{b} \qquad Z_{2} =
e^{iy/R\tilde{n}} u^{-b}
\end{equation}

$Z_{1}$ and $Z_{2}$ obey the following relation
\begin{equation}
       Z_{1}Z_{2} = e^{-2\pi i \theta_{0}} Z_{2} Z_{1}
\end{equation}

\noindent where
\begin{equation}
-\theta_{0} = \frac{a(-\theta) + b}{m(-\theta) + n}
\end{equation}

This relation is exactly the relation formed by $e^{ix'/R'}$ and
$e^{iy'/R'}$ where $[x',y'] = 2\pi i R^{\prime2}\theta_{0}$
(see equation (\ref{BCH})). For now, $R'$ is arbitrary.
The expansion (\ref{stdform}) is the Fourier expansion on
the noncommutative torus with no units of flux and non-commutativity
parameter $\theta_{0}$. A straightforward calculation shows
that any two theories related by the following transformation:
\be\label{Td}
	-\theta' = \frac{A(-\theta) + B}{C(-\theta) + D},\qquad
    \left(\begin{array}{c} n' \\ -m' \\ \end{array}\right) =
    \left(\begin{array}{cc} A & B\\ C & D \\ \end{array}\right)
    \left(\begin{array}{c} n \\ -m \\ \end{array}\right),
\ee
with $AD - BC = 1$, result in the same standard form. We will
determine the area of the dual torus at the end of this section.

In addition to defining the T-dual of the fields, we will also
have kinetic terms in our Lagrangian. Thus, we must define the T-dual
of the covariant derivative and the connection. Because
of the non-trivial boundary conditions, however, $A_{y}$ does not
transform in the adjoint so we cannot simply apply what we have done
above. For fields in the adjoint, the
covariant derivative is
\begin{equation}
       D_{\mu}\Psi = \partial_{\mu}\Psi + \left[A,\Psi\right]
\end{equation}

We wish to make a field for which we can easily define the T-dual.
   {}From $A_{y}$, we construct
\begin{equation}
       A'_{y} = A_{y} + \frac{imx}{2\pi R^{2}(n-m\theta)}
\end{equation}

\noindent This field transforms in the adjoint.
With this new field, we can write
\begin{equation}
       \label{coderiv}
       D_{y}\Psi = \frac{n}{n-m\theta} \partial_{y}\Psi +
            \left[A'_{y},\Psi\right]
\end{equation}

\noindent where we have transformed a commutator into a derivative as
in (\ref{com2der}).

Let us represent the operation
of T-duality by $T(\cdot)$. The new parameters will be as in
(\ref{Td}). For simplicity, we will work with the case $\gcd(n,m) =
1$. We define
the following operation closely related to T-duality:
\be\label{tstar}
        \Psi = \sum_{s,t\in\BZ} c_{s,t} Z_{1}^{s} Z_{2}^{-t},\qquad
	 T^{*}(\Psi) = \sum_{s,t\in\BZ} c_{s,t} Z_{1}^{\prime s}
	 Z_{2}^{\prime-t}
\ee

\noindent where the $Z$s and $Z'$s come from
the unprimed and primed theories which are dual.

For most fields, we have $T^{*} = T$, but for the
connection this is slightly modified. Instead, we have:
\be
	T(A_{y}) = T^{*}(A'_{y}) -
	\frac{im'x'}{2\pi R^{\prime 2}(n'-m'\theta')},\qquad
	T(A_{x}) = T^{*}(A_{x})
\ee

\noindent These have the correct boundary conditions for the T-dual
theory.

We need one more relation because $[T,\partial] \neq 0$:
\begin{eqnarray}
\partial_{\mu}\Psi &= &\sum_{s,t\in\BZ} c_{s,t} \batop{is/R(n-m\theta)}
{-it/Rn}Z_{1}^{s} Z_{2}^{-t} \\
T(\partial_{\mu}\Psi) &= &\sum_{s,t\in\BZ} c_{s,t} \batop{is/R(n-m\theta)}
{-it/Rn}Z_{1}^{\prime s} Z_{2}^{\prime-t} \\
\partial_{\mu}T(\Psi) &= &\sum_{s,t\in\BZ} c_{s,t}
\batop{is/R'(n'-m'\theta')}
        {-it/R'n'}Z_{1}^{\prime s} Z_{2}^{\prime-t}
\end{eqnarray}

\noindent where the top element in the brackets refers to the case $\mu = x$
and the bottom to the case $\mu = y$.

Therefore
\begin{equation}
T(\partial_{\mu}\Psi) = \frac{R'}{R}\batop{\frac{n'-m'\theta'}{n-m\theta}}
{\frac{n'}{n}} \partial_{\mu}T(\Psi)
\end{equation}

\noindent So,
\begin{eqnarray}
       \label{tcov}
       T(D_{\mu}\Psi) &= &\batop{1}{\frac{n}{n-m\theta}}
	 T(\partial_{\mu}\Psi) +
	 T(\left[A'_{\mu},\Psi\right]) \nonumber\\
	 &= &\frac{R'}{R}\left(\frac{n'-m'\theta'}{n-m\theta}\right)
  \batop{1}{
	 \frac{n'}{n'-m'\theta'}} \partial_{\mu}T(\Psi)
	 + \left[T(A'_{\mu}),T(\Psi)\right] \nonumber\\
	 &= & \batop{1}{\frac{n'}{n'-m'\theta'}} \partial_{\mu}T(\Psi) +
	 \left[T(A'_{\mu}),T(\Psi)\right] \nonumber\\
	 &= & D_{\mu}T(\Psi)
\end{eqnarray}

\noindent where we have set $R' = R(D-C\theta)$ to cancel the factor
on the second line. (Note that $\frac{n'-m'\theta'}{n-m\theta} =
\frac{1}{D-C\theta}$.)
This is consistent with the volume change $V' = V(D-C\theta)^{2}$
of \cite{SWNCG} and references therein.
Note that they use different lettering for
the $SL(2,\BZ)$ matrix which determines the T-duality transformation.
The transformation property of $g_{YM}$ can be determined from the
normalization of the integral, but we will not do so here.

\section{The T-dual of a Twist}\label{twists}
We now examine the effect of T-duality, as presented above, on a
field with twisted boundary conditions. We take the field $\phi$
to have boundary conditions:
\begin{eqnarray}
	\phi(x+2\pi R,y) &= &\Omega_{1}(x,y) \Psi(x,y) \Omega_{1}(x,y)^{-1},
	\nn \\
	\phi(x,y+2\pi R) &= &e^{2\pi i \alpha}\Omega_{2}(x,y) \Psi(x,y)
	\Omega_{2}(x,y)^{-1}
\end{eqnarray}

\noindent We wish to put this in a standard form. We define
\begin{equation}
       \phi' = e^{-i\alpha y/R}\phi
\end{equation}

\noindent This removes the twist so we can expand this field as
before:
\be
	\phi' = \sum_{s,t\in\BZ}c_{s,t}Z_{1}^{s}Z_{2}^{-t},
\qquad
	\phi = \sum_{s,t\in\BZ}c_{s,t}e^{i\alpha y/R}Z_{1}^{s}Z_{2}^{-t}
\ee

Now, we can tentatively define $T(\phi) = e^{i\alpha'y/R'}T^{*}(\phi')$.
This is not  precisely correct. For one, $\alpha'$ is still undetermined.
More importantly, there is a new behavior when we take the T-dual of a
product. To see this, we introduce:
\begin{equation}
       \psi = \sum_{u,v\in\BZ}d_{u,v}Z_{1}^{u}Z_{2}^{-v}
\end{equation}

\noindent Then
\begin{eqnarray}
	\psi\phi &= &\sum_{s,t,u,v\in\BZ} c_{s,t}d_{u,v}Z_{1}^{u}Z_{2}^{-v}
	e^{i\alpha y/R}Z_{1}^{s}Z_{2}^{-t}\nonumber\\
		 &= &\sum_{s,t,u,v\in\BZ} c_{s,t}d_{u,v}
		 e^{2\pi i\theta\alpha u /(n-m\theta)}e^{i\alpha y/R}
		 Z_{1}^{s}Z_{2}^{-t}Z_{1}^{u}Z_{2}^{-v}
\end{eqnarray}

\noindent However,
\begin{equation}
       T(\psi)T(\phi) = \sum_{s,t,u,v\in\BZ} c_{s,t}d_{u,v}
		 e^{2\pi i\theta'\alpha' u /(n'-m'\theta')}e^{i\alpha'y/R'}
		 Z_{1}^{\prime u}Z_{2}^{\prime-v}
		 Z_{1}^{\prime s}Z_{2}^{\prime-t}
\end{equation}

We can replicate the additional phase by translating one of the
fields. Thus,
\begin{equation}
      \label{Tshift}
	T(\psi\phi)(x,y) = T(\psi)(x+L,y)T(\phi)(x,y)
\end{equation}

\noindent where
\begin{equation}
       \label{ldef}
       L = 2\pi R'\left(\frac{\theta\alpha}{D-C\theta}
       - \theta'\alpha'\right)
\end{equation}

\noindent Because we also have $T(\phi\psi)(x,y) =
T(\phi)(x,y)T(\psi)(x,y)$ instead of the above, we define
\begin{equation}
       T(\phi) = \overleftarrow{E}_{L} e^{i\alpha'y/R'}T^{*}(\phi')
\end{equation}

\noindent where $\overleftarrow{E}_{L}$ is defined to obey:
\be
f(x)\overleftarrow{E}_{L} = \overleftarrow{E}_{L}f(x+L)
\ee

\noindent We will abuse the notation and
use $T(\phi)$ to represent the field without the shift operator which will be
explicitly shown in the multiplied fields as in (\ref{Tshift}).

We can now determine $\alpha'$ by examining how the covariant
derivative acts on these fields. The first difference with the
previous discussion comes from the derivative term:
\begin{equation}
       \label{twpartial}
       \partial_{y}\phi = i\frac{\alpha}{R}\phi + e^{i\alpha 
y/R}\partial_{y}\phi'
\end{equation}

\noindent The second difference comes from the reconstruction of the
normal form of the covariant derivative from the form
(\ref{coderiv}). Specifically, we have:
\begin{eqnarray}
       T([A'_{y},\phi]) &= &T(A'_{y})(x+L)T(\phi) -
            T(\phi)T(A'_{y}) \nonumber\\
         &= & T(A_{y})(x+L)T(\phi) - T(\phi)T(A_{y}) + \nn\\
	& &
            \frac{im'[x',T(\phi)]}{2\pi R^{\prime 2} (n'-m'\theta')}
            + \frac{im'LT(\phi)}{2\pi R^{\prime 2}(n'-m'\theta')}
\end{eqnarray}

Our goal is to have the covariant derivative satisfy (\ref{tcov}).
Expanding that, we obtain
\begin{eqnarray}
\lefteqn{
       \frac{n}{n-m\theta}\left(i\frac{\alpha}{R} T(\phi) +
       e^{i\alpha'y/R'}T(\partial_{y}\phi')\right) +
       \Big(T(A)(x+L)T(\phi) - T(\phi) T(A)(x)\Big)}\nn\\ &=&
       \frac{n'}{n'-m'\theta'}
    \left(i\frac{\alpha'}{R'} T(\phi) + 
e^{i\alpha'y/R'}\partial_{y}T(\phi')\right)
\\
    &+& \Big(T(A)(x+L)T(\phi) - T(\phi) T(A)(x)\Big)
     + \frac{im'LT(\phi)}{2\pi R^{\prime 2}(n'-m'\theta')}.
\nn
\end{eqnarray}

\noindent This implies the following condition:
\begin{equation}
       \frac{n\alpha}{R(n-m\theta)}= \frac{n'\alpha'}{R'(n'-m'\theta')}
       + \frac{m'L}{2\pi R^{\prime 2}(n'-m'\theta')}
\end{equation}

\noindent which is equivalent to:
\begin{equation}
     \label{leq}
     n\alpha = n'\alpha' + \frac{m'L}{2\pi R'}
\end{equation}

Now we have two equations in two variables that we can solve for $L$
and $\alpha'$. Solving (\ref{leq}) for $\alpha'$, we substitute into
(\ref{ldef}) giving
\begin{equation}
       L = 2\pi\alpha R' B \qquad \alpha' = D \alpha
\end{equation}

These remarkably simple answers are suggestive. If we examine the
transformation of a field of length $L$ and twist $\alpha$ into a
field of length $L'$ and twist $\alpha'$, equation (\ref{leq}) is
modified to
\begin{equation}
       n\alpha + \frac{mL}{2\pi R} = n'\alpha' + \frac{m'L'}{2\pi R'}
\end{equation}

\noindent and equation (\ref{ldef}) is modified to
\begin{equation}
       L' + 2\pi\theta'\alpha'R' = L + 2\pi\theta\alpha R
\end{equation}

Solving these gives
\begin{equation}
      \frac{L'}{2\pi R'} = \alpha B + A\frac{L}{2\pi R} \qquad \alpha' =
      \alpha D + \frac{L}{2\pi R}C
\end{equation}

\noindent or, in matrix form
\be
\left(\begin{array}{c}
       \frac{L'}{2\pi R'} \\ \alpha' \\ \end{array}\right)
=
\left(\begin{array}{cc}
	A & B \\ C & D \\ \end{array}\right)
\left(\begin{array}{c}
	 \frac{L}{2\pi R} \\ \alpha \\ \end{array}\right).
\ee

\noindent Note that both $\frac{L}{2\pi R}$ and $\alpha$ are parameters
that run from 0 to 1.

It is not hard to generalize the result for $\MT{d}$ with generic twists
$\a_1,\dots,\a_d$ and generic lengths $L_1,\dots,L_d$.
The T-duality group is $SO(d,d,\BZ)$ and 
$(\a_1,\dots,\a_d,\frac{L_1}{2\pi R_1},\dots,\frac{L_d}{2\pi R_d})$
transform as a vector in the representation $\rep{2d}$.
One can intuitively understand this result as follows.
The twists, $\a_i$, can be interpreted as a fractional momentum,
$p_i$, along the $i^{th}$ cycle. In noncommutative
geometry, T-duality exchanges momentum and electric flux
\cite{BraMor,KonSch}.
As we will see more clearly in the next subsection, the lengths
$L_i$ can be interpreted as dipole-lengths, and thus $L_i$ does
correspond to a fractional electric flux.

One might wonder about the zero modes of the dipole-fields.
When we compactify a scalar field with generic twisted boundary 
conditions the zero mode of the field disappears.
If we start with a local field with twisted boundary conditions,
we have seen that after a T-duality that acts as
$\Theta\rightarrow -1/\Theta$ we get a dipole-field with no
twist. At first sight it looks as if the dipole-field has
a zero mode. However, we have to recall that the T-dual theory
has some units of magnetic flux along $\MT{2}$.
In the presence of magnetic flux, the dipole-field has no zero
modes either. To see this, note that the dipole-field is charged
under the local gauge groups
 $U(1)_{(x_1,x_2)}\times U(1)_{(x_1+L_1,x_2)}$.
If we have $m$ units of magnetic flux
it is easy to see that the boundary conditions for 
$x_2\rightarrow x_2+2\pi R_2$  
include a phase $e^{\frac{2\pi i m L}{R_1}}$.

\section{Properties of Dipole Theories}\label{prop}
\def\wPhi{{\widetilde{\Phi}}}
\def\wstar{{\widetilde{\star}}}

We have seen in the previous section that noncommutative
field theories with scalars (or fermions) naturally lead us
to study field theories with dipoles.
The dipoles are described by a vector $L^\u$,
and they can be formulated on a commutative or noncommutative
spacetime.
In this section we will study these dipole theories.
We will see that they have features that resemble those of
noncommutative field theories, although they are much simpler.

\subsection{The modified $\star$-product}
We can construct a dipole theory by starting with a
field theory on a commutative or noncommutative space
and modify the $\star$-product (or regular product if
we are in the special case of  a commutative space).
To this end we designate a subset of the fields to
be ``dipole-fields'' and define the $\wstar$-product as follows.
If $\Phi$ is a dipole-field and $\Psi$ is an ordinary field
we set:
\be\label{dippro}
(\Phi \wstar \Psi)_{(x)} \equiv
\Phi(x)\star\Psi(x+L),\qquad
(\Psi \widetilde{\star} \Phi)_{(x)} \equiv
\Psi(x)\star\Phi(x).
\ee
More generally,
if $\Phi_1$ is a dipole-field with dipole-vector $L_1$
and $\Phi_2$ is a dipole-field with vector $L_2$,
we set:
\be
(\Phi_1 \wstar \Phi_2)_{(x)} \equiv
\Phi_1(x)\star\Phi_2(x+L_1),\qquad
(\Phi_2 \widetilde{\star} \Phi_1)_{(x)} \equiv
\Phi_2(x)\star\Phi_1(x+L_2).
\ee
Note that, in order for the $\wstar$-product to be associative,
the dipole-vector has to be additive. In other words, $\Phi_1\star\Phi_2$
should be defined to have dipole-vector $L_1 + L_2$.

If there are gauge fields, we define them to have dipole-vector zero.
The covariant derivative of a dipole-field becomes:
$$
(D_\u\Phi)_{(x)} \equiv
    \px{\u}\Phi +i A_\u\wstar \Phi -i \Phi\wstar A_\u
     = \px{\u}\Phi(x) +i A_\u(x)\star \Phi(x)
      -i \Phi(x)\star A_\u(x+L).
$$

\subsection{Seiberg-Witten map}
Seiberg and Witten described a map from the nonlocal noncommutative
gauge theory to a local theory with higher derivative interactions
\cite{SWNCG}.

Can we find a similar map that transforms the nonlocal dipole theories
to local theories with higher derivative interactions?
The map in this case is very simple.
Define:
$$
\Phi(x_1,\dots,x_d)
\equiv \wPhi(x_1,\dots,x_d)
    P e^{i L^\u \int_0^1 A_\u (x_1 + t L_1,\dots,x_d+t L_d) dt}.
$$
It is easy to see that if $\wPhi$ transforms in the adjoint
of the local gauge group at $(x_1,\dots,x_d)$ then
$\Phi$ transforms in the $(N,\bar{N})$ of
$U(N)_x\times U(N)_{x+ L}$.
We can also expand:
\bear
\left(D_\u\Phi\right)
    P e^{-i L^\u \int_0^1 A_\u (x_1 + t L_1,\dots,x_d+t L_d) dt}
&=&
D_\u\wPhi+ i L^\v \wPhi F_{\u\v} + \cdots
\label{swmap}
\eear
Using this map, we can write the first order, $O(L)$, correction
to the Lagrangian of a dipole scalar coupled  to a gauge field
as:
\bear
{\cal L} &=&
\frac{1}{4g^2} \tr{F_{\u\v} F^{\u\v}}
+\frac{1}{2}\tr{D_\u\wPhi D^\u\wPhi^\dagger}
+ L^\u\tr{J^\v F_{\u\v}} + O(L)^2,\nn\\
J^\v &\equiv&
    i D^\v\wPhi^\dagger \Phi
    -i \wPhi^\dagger D^\v\Phi.
\nn
\eear

\subsection{Rational dipoles}\label{rational}
If we compactify a field-theory on a noncommutative $\MT{2}$,
with noncommutativity given by $\theta^{ij} = \frac{p}{q}\epsilon^{ij}A$
(where $A$ is the area of $\MT{2}$ and $\frac{p}{q}$ is rational),
we can map the theory to a local field theory
on a $\MT{2}$ of area $A/q^2$ and some magnetic flux \cite{Bigatti}.
This also follows directly from T-duality as in \cite{SWNCG}.

Dipole theories have a similar property.
If we compactify a theory with dipoles of dipole-vector $L^\u$
on $\MS{1}$ of circumference $k L$ and such that $L^\u$ is in the
direction of $\MS{1}$, we can make it into a local theory on
$\MS{1}$ of radius $L$.
For example, take a gauge theory with gauge group
$U(N)$ or $SU(N)$ and a dipole scalar field of dipole-vector $L^\u$.
After compactification on $\MS{1}$ of circumference $k L$,
we can obtain a local gauge theory on $\MS{1}$ of radius $L$
and gauge group $U(N)^k$ or $SU(N)^k$ respectively.
The dipole-fields become local fields in the bifundamental
representation $(N_i, \overline{N}_{i+1})$. Here $i$ and $(i+1)$
refer to the $i^{th}$ and $(i+1)^{th}$ $U(N)$ or $SU(N)$ factors
in the chain $U(N)\times \cdots\times U(N)$, and if $i+1 = N+1$
we take $i+1\rightarrow 1$.
This theory has a $\BZ_k$ global symmetry of cyclically rotating
the chain. It is generated by $\sigma$ defined as taking
the $i^{th}$ $U(N)$ into the $(i+1)^{th}$ $U(N)$ and taking the
$(N_i, \overline{N}_{i+1})$ scalar into the
$(N_{i+1}, \overline{N}_{i+2})$ scalar.
The dipole-theory on $\MS{1}$ of radius $k L$ is equivalent to
this local quiver theory compactified on $\MS{1}$ of radius $L$
and with the boundary conditions $\phi(x+L) = \sigma\phi(x)$,
where $\sigma\phi$ denotes the action of $\sigma$ on any field $\phi$.

Note that a related limit of quiver theories appears in \cite{WitNGT}
in a different context.

Note also that in 3+1D the $U(N)^k$ quiver theories have a Landau
pole due to the $U(1)^k$ factors.
Combined with the results of the previous section,
this implies that noncommutative $U(N)$ gauge theories with matter
have a Landau pole, at least for certain twists.
In noncommutative Yang-Mills theory, the $U(1)$ factor does not
decouple \cite{SWNCG}.
If we set the noncommutativity to zero, the $SU(N)$ dipole-theories
are well-defined, and they do not have a Landau pole.

\subsection{S-duality}\label{sduality}

\def\btau{{\overline{\tau}}}
\def\cO{{{\cal O}}}
\def\wcO{{\widetilde{{\cal O}}}}

Consider 3+1D $\SUSY{4}$ $SU(N)$ SYM.
  In terms of $\SUSY{2}$ supersymmetry,
it contains a vector-multiplet and a hypermultiplet.
Now let us turn  on the dipole-moment for the hypermultiplet.
Namely, we replace the product of fields in the hypermultiplet
with the dipole product (\ref{dippro}) that depends on
the vector parameter $L^\u$.
What is the S-dual of this theory?

For spatial $L^\u$, the theory can be obtained by starting
with a 3+1D $SU(N)^k$ quiver-theory compactified on $\MS{1}$ of
radius $k L$ and taking the limit $k\rightarrow\infty$, as explained
in subsection (\ref{rational}).
At first sight, this would suggest
that the theory is S-dual to itself, since the
$\SUSY{2}$ quiver-theory is believed to be self-dual \cite{DouMoo}.

But now we face a puzzle!
For small $L^\u$,
using the relation (\ref{swmap}), we can write the dipole
theory as a deformation of $\SUSY{4}$ SYM of the form:
$$
{\cal L}\rightarrow {\cal L} +  L^\u {\cal O}_\u + \cdots,
$$
where ${\cal O}_\u$ is an operator of dimension 5 whose bosonic
part is:
\bear
{\cal O}_\u &=&
   {i\over {g_{YM}^2}}\sum_{a=1}^2
   \tr{\Phi_a^\dagger D^\v\Phi_a F_{\u\v}}
   +{i\over {g_{YM}^2}}\sum_{a=1}^2
   \tr{\Phi D^\v\Phi_a \com{\Phi^\dagger}{\Phi_a^\dagger}}
\nn\\ &&
   +{i\over {g_{YM}^2}}\tr{
   (\Phi_1 D^\v\Phi_2-\Phi_1 D^\v\Phi_2) \com{\Phi_1^\dagger}{\Phi_2^\dagger}}
    + {\mbox{c.c.}}
\nn
\eear
Here, $\Phi$ is the complex adjoint scalar of the vector-multiplet
and $\Phi_1,\Phi_2$ are the two scalars of the hypermultiplet.

For $L^\u=0$, $\SUSY{4}$ SYM is self-dual, with
$$
\tau\rightarrow
-\frac{1}{\tau},\qquad \tau\equiv
   {{4\pi i}\over {g_{YM}^2}}   +{{\theta}\over {2\pi}}.
$$
What happens to the operator ${\cal O}_\u$ under S-duality?
If the dipole theory is self-dual then,
by what has been said above, ${\cal O}_\u$ should be S-dual to itself.
But this statement is wrong! In fact, Intriligator has put forward a
list of conjectures about S-duals of various chiral primary operators
and their superconformal descendants in $\SUSY{4}$ SYM
\cite{IntBON}.\footnote{We are grateful to O. Aharony for reminding us
of this reference.}
The operator $\cO_\u$ is a descendant of the chiral
primary $\cO_3^{(IJK)} \equiv \tr{\Phi^{(I}\Phi^J\Phi^{K)}}$ where
$I,J,K=1\dots 6$ and $(IJK)$ means symmetrization with respect to
those indices.
{}From the appendix of \cite{IntBON} we see that there are two vector
operators of dimension 5 that are descendants of $\cO_3$.
They are both in the representation $\rep{15}$ of the R-symmetry
group which can be seen to be the same representation
as $\cO$. (The $\rep{15}$ can be generated by anti-symmetric
tensors $M_{IJ}$.)
The two operators are:
\bear
\delta^3\overline{\delta}\cO_3 &\rightarrow& \cO^{+}\equiv
\frac{1}{g_{YM}^3}
   \tr{F^{+}_{\u\v} \Phi^{\lbrack I} D^\v \Phi^{J\rbrack}} + \cdots,\nn\\
\overline{\delta}^3\delta\cO_3 &\rightarrow& \cO^{-}\equiv
   \frac{1}{g_{YM}^3}
   \tr{F^{-}_{\u\v} \Phi^{\lbrack I} D^\v \Phi^{J\rbrack}} + \cdots,\nn
\eear
Here we use the notation of \cite{IntBON} that $\delta$
and $\overline{\delta}$ are supersymmetry transformations,
$\lbrack IJ\rbrack$ represent anti-symmetrization with respect to
the $I,J$ indices, $F^{+}_{\u\v}$ and $F^{-}_{\u\v}$ stand for the
selfdual and anti-selfdual components of the field-strength $F_{\u\v}$,
and $(\cdots)$ stands for
terms that do not involve  the field-strength $F_{\u\v}$.
The conjecture of \cite{IntBON} is that under S-duality:
$$
\tau\rightarrow\frac{a\tau+b}{c\tau+d},
$$
the operators transform as:
\bear
\cO^{+} &\rightarrow&
   \left(\frac{c\btau+d}{c\tau+d}\right)^{1/4} \cO^{+},\nn\\
\cO^{-} &\rightarrow&
   \left(\frac{c\btau+d}{c\tau+d}\right)^{-1/4} \cO^{-},\nn
\eear
The operator we need  is:
$$
\cO_\u = g_{YM}^{} (\cO_\u^{+} + \cO_\u^{-}).
$$
We see that under S-duality it becomes a totally different operator
that contains the magnetic field-strength
$\widetilde{F}_{\u\v} \equiv \epsilon_{\u\v\sigma\rho}F^{\sigma\rho}$.
Thus, the S-dual deformation operator, $\wcO_\u$, contains time
derivatives. It is therefore likely that the S-dual theory,
for finite $L^\u$, is nonlocal in time.
This would be reminiscent of the nonlocality in time
  \cite{SSTi}-\cite{AhGoMe} that develops
after S-duality of spatially noncommutative SYM (NCSYM)
  \cite{GGS,GMMS}.
It would be interesting to investigate whether the dual theory
also contains string-like excitations similar to the duals of NCSYM
\cite{SSTii,GMMS}.

What was the flaw in the argument we presented at the beginning of
this subsection?
First, there is a factor of $k$ between the S-dual coupling
constant of the quiver-theory and the coupling constant of
$\SUSY{4}$ SYM. If we define $\tau$ in terms of the $\SUSY{4}$
SYM variables, then S-duality of the $SU(N)^k$ quiver-theory requires:
$$
k\tau\rightarrow -\frac{1}{k\tau}.
$$
If we take the limit $k\rightarrow\infty$ first, we see that
this duality replaces any finite $\tau$ with an infinitely strong
coupling constant $\tau=0$ and is therefore not the duality we
are seeking.
Moreover, it probably does not act locally in the sense that
it would not take local operators of the $\SUSY{4}$ theory
into local operators. That is,
after we unfold the circle of radius $L/k$
on which the $SU(N)^k$ quiver theory is defined
back to a circle of radius $L$ the notion of locality changes.

\subsection{Generalization to the $(2,0)$ theory}\label{circpol}
It has been argued in \cite{ABS,NekSch,Berk,SWNCG,GMSS} that
there exists a deformation of the $(2,0)$ theory that
depends on an anti-self-dual 3-form $\Th^{ijk}$ and,
after compactification on a small $\MS{1}$, the deformation
reduces to 4+1D SYM with noncommutativity parameter $\Th^{ij5}$.

One way to define this theory is to use a similar construction
as in \cite{DH}.
We start with $N$ M2-branes on $\MT{3}\times \MR{2,1}$.
The M2-branes span the directions of $\MR{2,1}$.
We include a 3-form C-field flux on $\MT{3}$ such that
the phase, $C\equiv C_{345} V_{345}$, is finite.
We then let the volume of the $\MT{3}$ shrink to zero:
$M_p^3 V_{345}\rightarrow 0$.
Since M-theory on $\MT{3}$ has an $SL(2,\BZ)$ subgroup
of the U-duality group that acts on  $\rho\equiv C + i V_{345}$ as
$$
\rho\rightarrow {{a\rho + b}\over {c\rho + d}},
$$
keeping the shape of $\MT{3}$ fixed (see for instance \cite{Ofer}),
the arguments of \cite{DH} seem to carry over.
This presumably describes the OM
on $\MT{3}$ with the same shape and with area $V$
such that $\Th^{345} = \Th^{012} = C V$.

   {}From this definition it is obvious that the theory has a U-duality
$SL(2,\BZ)$ group that acts nonlinearly on $\Th^{345}$.
We can ask the same question as before, namely what happens
if we compactify with an R-symmetry twist on the scalar fields
along the $5^{th}$ direction and apply U-duality.

   The answer should be a 5+1D theory that depends on an anti-symmetric
tensor $L^{\u\v}$ (in this case only $L^{34}$ is expected to be nonzero).
Let us take the R-symmetry twist to be such that it breaks half of
the supersymmetries. Note that since we break Lorenz invariance explicitly
we can end up with a theory with 8 supersymmetries in 5+1D.
If we expand in $L^{\u\v}$, we expect the leading order to be
a dimension-8 operator ${\cal O}_{\u\v}$.

Let us take the case $N=1$ (a single M5-brane), and
let us denote the 5 scalar fields of a free tensor multiplet
by $\Phi^I$ ($I=1\dots 5$). Let $\phi_a$ ($a=1,2$) be the complex
fields:
$$
\phi_1 \equiv \Phi_2 + i \Phi_3,\qquad \phi_2\equiv \Phi_4 + i\Phi_5.
$$
We expect the bosonic part of ${\cal O}_{\u\v}$ to be:
$$
{\cal O}_{\u\v} = i\sum_{a=1}^2
(\phi_a^\dagger\qx{\sigma}\phi^a
     -\qx{\sigma}\phi_a^\dagger\phi^a) H_{\u\v\sigma}.
$$
Here $H_{\u\v\sigma}$ is the 3-form field-strength of the free-tensor
multiplet and $\phi^a$ are the 2 complex scalar fields defined above.

If we turn on, for example, only $L^{34}$, then
the particles that the fields $\phi_a$ describe seem to be,
instead of dipoles,
two dimensional surfaces of constant area in the $3-4$ plane.
The contour is charged under the field $H_{\u\v\sigma}$.\footnote{
As this work was in progress, we found out about related ideas
on the ``noncommutative'' $(2,0)$ theory \cite{BerkTA}.
In this work, somewhat similar ideas seem to have been reached from a
different path. We are grateful to M. Berkooz for
discussing his ideas with us.}
These contours seem to be dynamical.
It would be interesting to see if an action can be written
for these contours.


\section{Discussion}\label{disc}

Noncommutative field theories have proven to be an exciting nonlocal
generalization
of ordinary field-theories that provide a testing ground for stringy
phenomena. The generalization to the ``noncommutative'' $(2,0)$
theory at rank $N=1$ is especially intriguing
if it can provide insight into the $(2,0)$ theory itself.

It is therefore interesting to discover that there exist simplified
versions of these theories.
In this paper we have argued that studying noncommutative field
theories with twisted boundary conditions naturally leads us to consider
dipole theories. As we have demonstrated in section (\ref{prop}),
these theories have properties that mimic those of noncommutative
field theories, but in a very simplified fashion.
It therefore seems worthwhile to study the other nonlocal phenomena
of noncommutative gauge theories in the context of dipole theories.
Hopefully, the simplified form of dipole theories might shed
more light on the nonlocal phenomena.
For example:
\begin{enumerate}
\item
Timelike noncommutativity \cite{SSTi}
appears to require inclusion of
stringy degrees of freedom \cite{SSTii}-\cite{AhGoMe},\cite{GMMS}.
This is in part
motivated by the study of strings in critical electric fields
\cite{HVerl,GuKlPo}.
It would be interesting to study dipole theories with timelike
dipole-vectors.

\item
Theories with
lightlike noncommutativity were studied in \cite{AhGoMe,AlOzRu}.
It would be interesting to extend the study to lightlike dipole-vectors.

\item
The arguments in subsection (\ref{sduality}) suggest that
the S-dual of dipole theories might be nonlocal in time.
It would be interesting to verify this and perhaps describe
the S-dual theory more explicitly.
It might be reminiscent of the NCOS theories \cite{SSTii,GMMS}.

\item
It is very interesting to explore the extensions to the $(2,0)$ theories
discussed in subsection (\ref{circpol}).
We have suggested that they involve some kind of generalization
of dipoles of fixed length to 2D contours bounding a fixed area.
   Perhaps it would be possible
to quantize (first quantization) the degrees of freedom in the boundary
of these contours. (See also the related ideas in \cite{BerkTA}.)

\item
In \cite{CDGG} various nonlocal theories with the nonlocality
being characterized by a vector have been suggested.
They were constructed by placing D-branes in backgrounds that
carry $B^{NSNS}$ field-strength. These backgrounds were obtained
by starting with a Taub-NUT space and setting the boundary conditions
such that a component of the NSNS
$B$-field, with one index along the Taub-NUT circle
and one index parallel to the D-brane, is a nonzero
constant at infinity.
It would be interesting to understand the relation between those theories
and the dipole theories.



\item
In \cite{CGK,CGKM}, the T-dual of an R-symmetry twist in the context
of little-string theories \cite{SeiVBR} was studied. It would be
interesting to see if any insight on these mysterious T-dual
twists can be gained from the dipole theories.
\end{enumerate}

\section*{Acknowledgments}
We are grateful to O. Aharony, M. Berkooz, Govindan Rajesh, M. Gremm
and S. Sethi for discussions.
This research is supported by NSF grant number PHY-9802498.


\def\np#1#2#3{{\it Nucl.\ Phys.} {\bf B#1} (#2) #3}
\def\pl#1#2#3{{\it Phys.\ Lett.} {\bf B#1} (#2) #3}
\def\physrev#1#2#3{{\it Phys.\ Rev.\ Lett.} {\bf #1} (#2) #3}
\def\prd#1#2#3{{\it Phys.\ Rev.} {\bf D#1} (#2) #3}
\def\ap#1#2#3{{\it Ann.\ Phys.} {\bf #1} (#2) #3}
\def\ppt#1#2#3{{\it Phys.\ Rep.} {\bf #1} (#2) #3}
\def\rmp#1#2#3{{\it Rev.\ Mod.\ Phys.} {\bf #1} (#2) #3}
\def\cmp#1#2#3{{\it Comm.\ Math.\ Phys.} {\bf #1} (#2) #3}
\def\mpla#1#2#3{{\it Mod.\ Phys.\ Lett.} {\bf #1} (#2) #3}
\def\jhep#1#2#3{{\it JHEP.} {\bf #1} (#2) #3}
\def\atmp#1#2#3{{\it Adv.\ Theor.\ Math.\ Phys.} {\bf #1} (#2) #3}
\def\jgp#1#2#3{{\it J.\ Geom.\ Phys.} {\bf #1} (#2) #3}
\def\cqg#1#2#3{{\it Class.\ Quant.\ Grav.} {\bf #1} (#2) #3}
\def\hepth#1{{\it hep-th/{#1}}}


\end{document}